\documentclass[12pt]{article}
\usepackage{a4}
\usepackage{amssymb,amsmath,amsfonts}
\usepackage{epsfig}
\def\bfit#1{\textbf{\textit{#1}}}

\begin{document}

\title{On the reduction of hypercubic lattice artifacts}
\date{\today}
\author{ F.~de~Soto$^{a,b}$ and C.~Roiesnel$^c$ }
\par \maketitle
\begin{center}
$^a$ Dpto. de Sistemas F\'{\i}sicos, Qu\'{\i}micos y Naturales,\\
Universidad Pablo de Olavide, Ctra. Utrera Km 1, 41013 Sevilla, Spain.\\
$^b$ Laboratoire de Physique Subatomique et Cosmologie, IN2P3, \\
53 av. des Martyrs, 38026 Grenoble, France. \\
$^c$ Centre de Physique Th\'eorique, Ecole Polytechnique, CNRS,\\
91128 Palaiseau, France.
\end{center}

\begin{abstract}
  This note presents a comparative study of various options to reduce
  the errors coming from the discretization of a Quantum Field Theory
  in a lattice with hypercubic symmetry. We show that it is possible
  to perform an extrapolation towards the continuum which is able to
  eliminate systematically the artifacts which break the $O(4)$
  symmetry.
\end{abstract}
\vfill
\begin{flushleft}
CPHT RR 016.0307\\
\end{flushleft}

\enlargethispage{0.5cm}

\newpage

\section{Introduction}
\label{sec:intro}

The problem of restoration of rotational invariance was the focus of
much work in the early days of numerical simulations of lattice gauge
theories, which were performed on very small lattices. Most noteworthy
were the attempts to find alternative discretizations which would
approach the continuum limit more rapidly than the simple hypercubic
lattice. One line of attack \cite{CEL82} was to discretize gauge
theories on the most symmetric of all four-dimensional lattices, the
four-dimensional body-centered hypercubic lattice, whose point
symmetry group is three times as large as the hypercubic group.
Another angle of investigation worth mentioning was to formulate gauge
theories on random lattices \cite{LEE82}. The interest in these
alternate formulations faded away in subsequent years, first because
of their inherent complications, but mainly when it was realized that
rotational invariance was in fact restored within statistical errors
at larger distances on the hypercubic lattice.

However, the treatment of discretization errors in numerical
simulations of a lattice gauge theory can remain a vexing problem
in some data analyses. Indeed, the signal of some lattice
observables, such as the two-point Green functions in momentum
space, has become so good that the systematic errors become very
much larger than the statistical errors. A general method, which
we call the H4 method, has been devised quite some time ago
\cite{LPTX99,LPTX00} to eliminate hypercubic artifacts from the
gluon two-point functions and extrapolate the lattice data towards
the continuum. This extrapolation is crucial to succeed in a
quantitative description, at least in the ultraviolet regime. Such
a method, despite its success in describing other two-point
functions as well, as the fermion \cite{LPT03} or the ghost
\cite{LPTX05} propagators, has not been widely adopted. Indeed,
most other studies of the lattice two-point functions are still
using phenomenological recipes \cite{ADELAIDE} which only allow
for a qualitative description of the data, since it is usually not
possible to make quantitative fits with a reasonable chisquare.

The purpose of this note is threefold. First we want to gather some
pieces about the H4 technique which are scattered in various sections
of previous publications and which may have been overlooked. Our
second objective is to stress, on a simple controllable model, that
the H4 method can be systematically improved, contrarily to the
empirical methods, when the statistical errors decrease. Our last goal
is to point out the general applicability of the method, not only to
those scalar form factors in momentum space which depend on a single
invariant, but also to various other lattice observables.

The plan of the paper is as follows. In the next section we recall
the general technique of hypercubic extrapolations towards the
continuum of any lattice scalar form factor depending upon a
single momentum. In the following section we show that a simple
model, a free real scalar field in four dimensions, can be used as
a testbed for the hypercubic extrapolations. Then we make a
detailed comparison of the different strategies to eliminate the
hypercubic lattice artifacts. The concluding section is devoted to
recommendations about the best usage of the H4 extrapolation
method. We also outline some straightforward generalizations.

\section{Hypercubic artifacts}
\label{sec:artifact}

Any form factor $F_L(p)$ which is a scalar invariant on the lattice,
is invariant along the orbit $O(p)$ generated by the action of the
isometry group $H(4)$ of hypercubic lattices on the discrete momentum
$p\equiv\frac{2\pi}{La}\times(n_{1},n_{2},n_{3},n_{4})$ where the
$n_{\mu}$'s are integers, $L$ is the lattice size and $a$ the lattice
spacing.  The general structure of polynomial invariants under a
finite group is known from group-invariant theory \cite{WEYL}. In
particular, it can be shown that any polynomial function of $p$ which
is invariant under the action of $H(4)$ is a polynomial function of
the 4 invariants
\begin{align}
p^{[n]} \equiv \sum_{\mu}p_{\mu}^{n},\quad n = 2, 4, 6, 8
\end{align}
which index the set of orbits. The appendix contains an elementary
derivation.

It is thus possible to use these 4 invariants to average the form
factor over the orbits of $H(4)$ to increase the statistical accuracy:
\begin{align}
  F_L(p) \equiv F_{L}(p^{[2]},p^{[4]},p^{[6]},p^{[8]}) =
  \frac{1}{\|O(p)\|} \sum_{p\in O(p)} F_L(p)
\end{align}
where $\|O(p)\|$ is the cardinal number of the orbit $O(p)$.

The orbits of the continuum isometry group $O(4)$ are of course
labeled by the single invariant $p^{[2]}\equiv p^2$, and lattice
momenta which belong to the same orbit of $O(4)$ do not belong in
general to the same orbit of $H(4)$. For instance, as soon as
$n^2\equiv\sum_{\mu=1}^{4}n^2_{\mu}=4$ in integer lattice units,
the $O(4)$ orbit splits into two distinct H(4) orbits, those of
the vectors $(2,0,0,0)$ and $(1,1,1,1)$ respectively. Therefore we
can distinguish two kinds of lattice artifacts, those which depend
only upon the invariant $p^{2}$, and which produce the scaling
violations, and those which depend also upon the higher-order
invariants $p^{[n]}$ ($n= 4,6,8$) and which we call hypercubic
artifacts. When the difference between the values of $F_L(p)$
along one orbit of $O(4)$ become larger than the statistical
errors, one needs at least to reduce the hypercubic artifacts from
the lattice data before attempting any quantitative analysis.

The treatment of these discretization artifacts can be inferred
from lattice perturbation theory, as Green functions will depend
on some lattice momentum~\footnote{Depending on the discretization
scheme, it will be $\widehat{p}_\mu$ or
$\overline{p}_\mu=\frac{1}{a}\sin{a p_\mu}$, etc.}
\begin{equation}
\widehat{p}_\mu\equiv \frac{2}{a} \sin\left(\frac{ap_\mu}{2}\right) \
\end{equation}
instead of the continuum one, $p_\mu=\frac{2\pi}{La}n_\mu$. By
developing the lattice momentum $\widehat{p}^2\equiv \sum_\mu
\widehat{p}_\mu^2$ in terms of the lattice spacing $a$, one gets:
\begin{equation}
\widehat{p}^2 \approx\ p^2 - \frac{a^2}{12} p^{[4]} + \frac{a^4}{360}
p^{[6]} - \frac{a^6}{20160} p^{[8]} + \cdots \label{hatkdev}
\end{equation}
and thus, the lattice momentum differs from the "continuum" one by
discretization artifacts that are proportional to the invariants
$p^{[4]}$ (of order $a^2$), $p^{[6]}$ (order $a^4$), etc.

The strategies to minimize the hypercubic artifacts are based on the
fact these artifacts depend on the non O(4) invariants, $p^{[4]}$,
$p^{[6]}$, etc. and thus reducing $p^{[4]}$ would also reduce the
artifacts. For example, the improved restoration of the rotational
symmetry on the four-dimensional body-centered hypercubic lattice can
be analyzed in terms of the primitive invariant $p^{[4]}$ \cite{NEU87}
\footnote{We thank Ph.~de~Forcrand for pointing out this reference to
  us.}. These strategies fall into three general groups:

\begin{itemize}
\item The simplest one is just to keep only the H(4) orbits which
minimizes  $p^{[4]}$ along each O(4) orbit. As they lay near the
diagonal, a more efficient prescription \cite{ADELAIDE} is to
impose a "cylindrical" cut on the values of $p$, keeping only
those that are within a prescribed distance of the diagonal. This
completely empirical recipe has been widely adopted in the
literature and we shall refer to it in the sequel as the
``democratic'' method. The main drawbacks are that the information
for most of the momenta is lost (for moderate lattices only a
small fraction of the momenta is kept) and that although $p^{[4]}$
is small for the orbits kept, it is not null, and therefore the
systematic errors are still present.

\item The other methods try to fully eliminate the contribution of
  $p^{[4]}$, etc. and we will generically refer to them as the H4
  methods. By analogy with the free lattice propagators, it is natural
  to make the hypothesis that the lattice form factor is a smooth
  function of the discrete invariants $p^{[n]}$, $n\geq 4$, near the
  continuum limit,
\begin{align}
  \label{eq:invariants}
  \begin{split}
  F_{L}(p^{2},p^{[4]},p^{[6]},p^{[8]}) &\approx
  F_{L}(p^{2},0,0,0) + p^{[4]}\frac{\partial F_{L}}
  {\partial p^{[4]}}(p^{2},0,0,0) + \\
  &\quad p^{[6]}\frac{\partial F_{L}}{\partial p^{[6]}}(p^{2},0,0,0)
  + (p^{[4]})^2\frac{\partial^2 F_{L}}{\partial^2 p^{[4]}}(p^{2},0,0,0)
  + \cdots
  \end{split}
\end{align}
and $F_{L}(p^{2},0,0,0)$ is nothing but the form factor of the
continuum in a finite volume, up to lattice artifacts which do not
break $O(4)$ invariance and which are true scaling violations.  We
emphasize that we are merely conjecturing that the restoration of
rotational invariance is smooth when taking the continuum limit at
fixed $p^2$. When several orbits exist with the same $p^2$, the
simplest method \cite{LPTX99} to reduce the hypercubic artifacts is to
extrapolate the lattice data towards $F_{L}(p^{2},0,0,0)$ by making a
linear regression at fixed $p^{2}$ with respect to the invariant
$p^{[4]}$ (note that the contributions of other invariants are of
higher order in the lattice spacing).

Obviously this method only applies to the O(4) orbits with more than
one H(4) orbit. If one wants to include in the data analysis the
values of $p^2$ with a single H(4) orbit, one must interpolate the
slopes extracted from \eqref{eq:invariants}. This interpolation can be
done either numerically or by assuming a functional dependence of the
slope with respect to $p^{2}$ based, for example, on dimensional
arguments \cite{LPTX00}. For instance, for a massive scalar lattice
two-point function, the simplest ansatz would be to assume that the
slope has the same leading behavior as for a free lattice propagator:
\begin{align}
  \label{eq:slope}
  \frac{\partial F_{L}}
  {\partial p^{[4]}}(p^{2},0,0,0) &=
 \frac{a^2}{\left(p^{2}+m^2\right)^{2}}\left( c_{1}+ c_{2}a^2p^{2}\right)
\end{align}
The range of validity of the method can be checked a posteriori
from the smoothness of the extrapolated data with respect to
$p^2$. The quality of the two-parameter fit to the slopes, and the
extension of the fitting window in $p^{2}$, supplies still another
independent check of the validity of the extrapolations, although
the inclusion of $O(4)$-invariant lattice spacing corrections is
usually required to get fits with a reasonable $\chi^{2}$.

This strategy based on independent extrapolations for each value
of $p^2$ will be referred to as the local H4 method.

\item The number of distinct orbits at each $p^{2}$ --in physical
  units-- increases with the lattice size and, eventually, a linear
  extrapolation limited to the single invariant $p^{[4]}$ breaks down.
  But, by the same token, it becomes possible to improve the local H4
  method by performing a linear regression at fixed $p^2$ in the
  higher-order invariants as well.  Therefore, when the lattice size
  increases, the H4 technique provides a systematic way to include
  higher-order invariants and to extend the range of validity of the
  extrapolation towards the continuum. For those $p^2$ which do not
  have enough orbits to perform the extrapolation, it is still
  possible to make use of all available physical information in the
  modelling of the functional derivatives appearing in
  \eqref{eq:invariants} and to perform an interpolation.

  An alternative strategy is based on the fact that the functional
  derivatives which appear in \eqref{eq:invariants} are functions of
  $p^2$ only. These functions can be represented by a Taylor
  development in their domain of analyticity, or, more conveniently,
  by a Laurent series, as it does not assume analyticity and makes
  appear all the terms allowed by dimensional arguments. Moreover, it
  is always possible to use polynomial approximation theory and expand
  the functional derivatives in terms of, e.g., Chebyshev polynomials
  or in a fourier series, etc.

In any case, these linear expansions allow to perform the
continuum extrapolation through a global linear fit of the
parameters for all values of $p^2$ inside a window at once.  Such
a strategy has been developed for the analysis of the quark
propagator \cite{LPT03} and we shall refer to it as the global H4
method. The global H4 extrapolation is simple to implement since
the numerical task amounts to solving a linear system.  It
provides a systematic way to extend the range of validity of the
extrapolation towards the continuum, not only for large lattices
(where the inclusion of $O(a^4)$ and even $O(a^6)$ discretization
errors becomes possible) but also for small lattices (where the
local H4 method for $O(a^2)$ errors is inefficient due to the
small number of orbits), by using in the fit all available lattice
data points.

\end{itemize}

\section{The free scalar field}

In order to analyze a model simple enough to provide a complete
control of the hypercubic errors in four dimensions, we have
chosen a free real scalar field, whose dynamics is given by the
lagrangian:
\begin{align}
\mathcal{L} \ =\ \frac{1}{2} m^2 \phi(x) \phi(x) + \frac{1}{2}
\partial_\mu\phi(x) \partial^\mu\phi(x)
\label{lcon}
\end{align}
The naive discretization of \eqref{lcon} leads to the lattice action:
\begin{align}
S \ =\ \frac{a^4}{2} \sum_x \left\{m^2 \phi_x^2 + \sum_{\mu=1}^{4}
(\nabla_\mu \phi_x)^2 \right\}
\end{align}
where $\nabla_{\mu}$ is the forward lattice derivative, or in momentum
space,
\begin{align}
S \ =\ \frac{a^4}{2} \sum_p \left(m^2 + \widehat{p}^2\right)
|\widetilde{\phi}_p|^2
\end{align}
where $p$ is the discrete lattice momentum. Therefore, the field
$\widetilde{\phi}_p$ can be produced by means of a gaussian
sampling with standard deviation $\sqrt{m^2 +
  \widehat{p}^2}$. As this is a cheap lattice calculation, we
can go to rather big volumes, up to $64^4$ in this work, and we can
generate a high number of fully decorrelated configurations.  In order
to study the effect of statistics over the results, averages will be
made over ensembles of $100$ till $1000$ configurations.

\begin{figure}[!ht]
\begin{center}
\epsfxsize10cm\epsffile{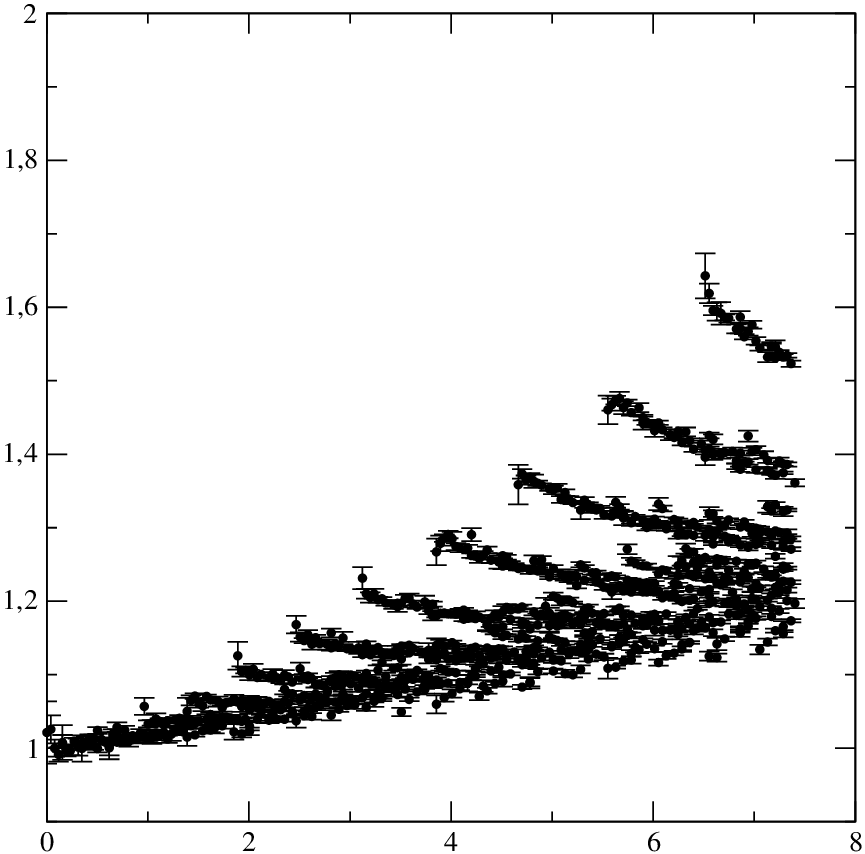}
\end{center}
\caption{\small{\it Raw dressing function
$\Delta_L(p)/\Delta_0(p)$ as a
    function of $p^2/m^2$ for a $32^4$ lattice and $am=1$ from a
    sample size of 1000 configurations.}}
\label{bonefish}
\end{figure}

This lattice model is of course solvable, and the propagator reads:
\begin{align}
\label{plat} \Delta_L (p)\ =\ \frac{1}{\widehat{p}^2+m^2}
\end{align}
The lattice artifacts are exactly computable by expanding
$\widehat{p}^2$ in terms of the $H(4)$ invariants introduced in the
previous section  and plugging the development (\ref{hatkdev})
into \eqref{plat},
\begin{align}
\label{devprop}
  \Delta_L(p^2,p^{[4]},p^{[6]},p^{[8]}) &\approx \frac{1}{p^2+m^2}
\ +\  a^2 \left\{ \frac{1}{12} \frac{p^{[4]}}{(p^2+m^2)^2}
\right\}
\nonumber \\
&+ a^4 \left\{ \frac{1}{72} \frac{{p^{[4]}}^2}{(p^2+m^2)^3} -
            \frac{2}{8!}\frac{p^{[6]}}{(p^2+m^2)^2} \right\}+\cdots
\end{align}
and the continuum propagator $\Delta_0(p)$ is indeed recovered
\bfit{smoothly} in the limit $a\to0$.  But as long as we are
working at finite lattice spacing, there will be corrections in
$a^2$, $a^4$, etc. that are not at all negligible, as can be
appreciated in figure \ref{bonefish} which plots the ratio
$\Delta_L(p)/\Delta_0(p)$ for a $32^4$ lattice.

One could wonder whether such a model is really useful since the
lattice artifacts are exactly known. For instance one can recover the
continuum propagator from the lattice propagator by merely plotting
the lattice data as a function of $\widehat{p}^2$ rather than $p^2$!
However this simple recipe is no longer applicable to an interacting
theory where the lattice two-point functions \bfit{do depend} upon the
independent variables $\widehat{p}^{[n]} =
\sum_{\mu}(\widehat{p}_{\mu})^{n},\ n=4,6,8$ (as illustrated in figure
1 of reference \cite{LPTX99}). And there is no systematic way to
separate out cleanly the effect of these additional variables because
$\widehat{p}^2$ is not an $O(4)$ invariant. Indeed, because
$\widehat{p}^2$ takes on different values on every $H(4)$ orbit, there
is only one data point per value of $\widehat{p}^2$ and the H4 method,
either local or global, is not appropriate for the choice of momentum
variable $\widehat{p}$.

However one should exercise special attention at using this model
without the information provided by expression \eqref{devprop}
(except of course the smoothness assumption in the $H(4)$
invariants $p^{[n]}$).  Under this proviso, the model can serve as
a bench test of the different approaches to eliminate hypercubic
artifacts. In particular {\it we will not use}
Eq.\,\eqref{eq:slope}.

The model has one mass parameter $m$ which fixes the scale. We
will study the worst-case scenario where $m$ cannot be neglected
with respect to $p$ when the lattice artifacts are
large\footnote{As
  $p=\frac{2\pi}{La}n$, with $n=0,\cdots,L/2$, a suitable value is $am=1$.}.

\textit{The case of QCD is, in fact,  simpler, as long as
$\Lambda_{QCD}$ and quark masses are negligible in comparison to
the momentum scale, which would correspond to the case $am\ll 1$.
Then, by dimensional arguments, the artifacts can be modeled at
least in the ultraviolet regime, as proposed in \cite{LPTX00} and
\cite{LPT03}.}

\section{Comparative study of H4 extrapolations}
\label{sec:H4}

We will now use a free scalar field with $am=1$ to compare the
different strategies to extract the continuum behavior from the
lattice data. We will use lattice units and set $a\equiv 1$
throughout this section. We restrict ourselves to one or two
representative methods within each strategy:
\begin{itemize}
\item The democratic method with a cylindrical cut selecting out the
  orbits that are within a distance of 2 lattice units from the diagonal
  $(1,1,1,1)$.
\item The local H4 method with independent extrapolations up to
  $\mathcal{O}(a^2)$ artifacts for every $p^2$ with several orbits
  within the window $n^2>5$ ($p=\frac{2\pi}{L}n$) up to some
  $n^2_{max}$:
  \[\Delta_L(p^2,p^{[4]},p^{[6]},p^{[8]}) = \Delta_L(p^2,0,0,0,0) +
  c(p^2)p^{[4]} \] The slopes $c(p^2)$ are then fitted with the following
  functional form
  \begin{align}
  c(p^2) = \frac{c_{-1}}{p^2} + c_0 + c_1p^2
\end{align}
which is used to extrapolate the points with only one orbit inside the
window $]5,n^2_{max}]$.
\item The global H4 methods with the coefficients of the artifacts up
  to $\mathcal{O}(a^2)$ or up to $\mathcal{O}(a^4)$ chosen as a
  Laurent series:
  \begin{align}
    \label{eq:global}
    \nonumber
    \Delta_L(p^2,p^{[4]},p^{[6]},p^{[8]}) &= \Delta_L(p^2,0,0,0,0) +
  f_1(p^2)p^{[4]} + f_2(p^2)p^{[6]} + f_3(p^2)(p^{[4]})^2 \\
  f_n(p^2) &= \sum_{i=-1}^{1} c_{i,n}(p^2)^{-i}\,,\quad n=1,2,3
  \end{align}
  With such a choice, a global fit within the window $]5,n^2_{max}]$
  amounts to solving a linear system of respectively $n^2_{max}-2$ and
  $n^2_{max}+4$ equations~\footnote{Those variables correspond
  respectively to the extrapolated propagators, $\Delta_L(p^2,0,0,0,0)$,
  and the 3 coefficients of each Laurent series.}.
\end{itemize}
Notice that we do not use the knowledge of the mass, $m=1$, in
both the local H4 method and the global H4 method, neither
directly nor indirectly (by introducing a mass scale as a
parameter). Our purpose is to stress the H4 extrapolation methods
to their limits. In practice, of course, all the physical
information can be used in order to improve the elimination of the
discretization artifacts.

\begin{figure}[!ht]
\begin{center}
\epsfxsize10cm\epsffile{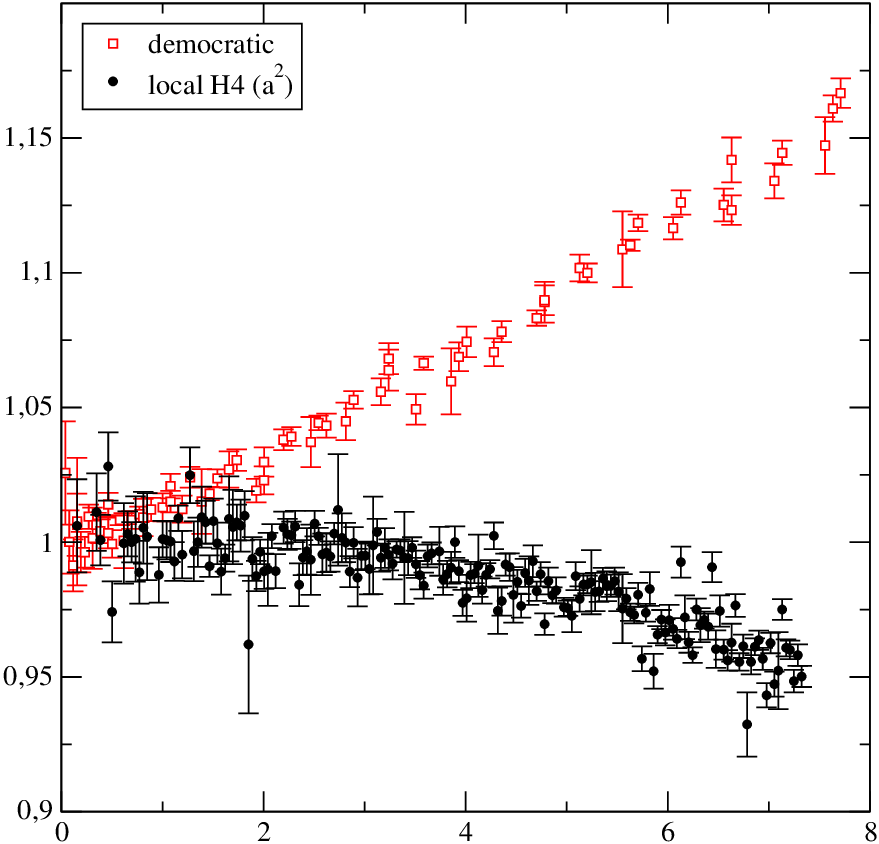}
\end{center}
\caption{\small{\it Comparison of the extrapolated dressing function
    $\Delta_E(p^2)/\Delta_0(p^2)$ as a function of $p^2$ on a
    $32^4$ lattice ($a=m=1)$, between the democratic method (open
    squares) and the local H4 method (black circles) -
    {\bf 1000 configurations}.}}
\label{fig:dem_vs_local}
\end{figure}

In figure~\ref{fig:dem_vs_local} the extrapolated dressing
functions $\Delta_E(p^2)/\Delta_0(p^2)$, with the notation
$\Delta_E(p^2)\equiv \Delta_L(p^2,0,0,0)$, of the democratic
method and of the local H4 method (with $p^2_{max}=3\pi^2/4$), are
compared for 1000 configurations generated on a $32^4$ lattice. It
can be seen that the dressing function of the democratic method
deviates very early from unity whereas the dressing function of
the local H4 method is pretty consistent with unity within
statistical errors for $p^2$ up to $\approx\pi^2/4$.

\begin{figure}[!ht]
\begin{center}
\epsfxsize10cm\epsffile{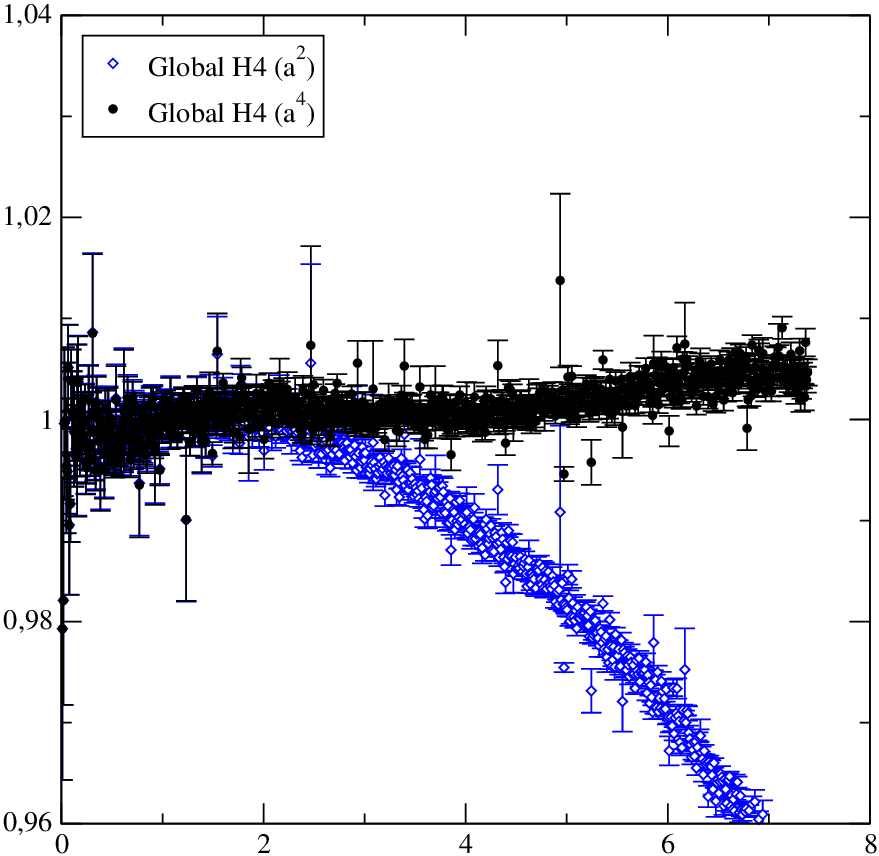}
\end{center}
\caption{\small{\it Comparison of the extrapolated dressing function
    $\Delta_E(p^2)/\Delta_0(p^2)$ as a function of $p^2$ on a $64^4$
    lattice ($a=m=1$), between the global methods with
    $\mathcal{O}(a^2)$ artifacts (open losanges) and $\mathcal{O}(a^4)$
    (black circles) - {\bf 1000 configurations}.}}
\label{fig:global}
\end{figure}

Figure~\ref{fig:global} compares the extrapolated dressing
functions of the global H4 methods, with respectively up to
$\mathcal{O}(a^2)$ and up to $\mathcal{O}(a^4)$ artifacts (again
with $p^2_{max}=3\pi^2/4$), for 1000 configurations generated on a
$64^4$ lattice. The global H4 method up to $\mathcal{O}(a^2)$
performs roughly as the local H4 method.  The global H4 method
which takes into account $\mathcal{O}(a^4)$ artifacts is able to
reproduce the continuum dressing function within statistical
errors for $p^2$ up to $\approx \pi^2/2$.

It is possible to put these qualitative observations on a more
quantitative basis, and show precisely the effect of both the lattice
size and the sample size on each extrapolation method. Since
all components of a free scalar field in momentum space are
independent gaussian variables, the statistical distribution of the
quantity
\begin{align}
\label{eq:chi2}
  \chi^2 = \sum_{p^2=1}^{p^2_{max}}
  \left(\frac{\Delta_E(p^2) - \Delta_0(p^2)}{\delta\Delta_E(p^2)} \right)^2
\end{align}
should follow exactly the chisquare law for $n^2_{max}$
independent variables, if the systematic errors of an
extrapolation method are indeed smaller than the statistical
errors. The criterion is exact for the democratic and local H4
methods which produce independent extrapolated values.
Extrapolations by the global H4 method are correlated and one must
include the full covariance matrix of the fit in the definition of
the chisquare:
\begin{equation}
\chi^2 = \sum_{p^2=1}^{p^2_{max}} \sum_{q^2=1}^{p^2_{max}}
(\Delta_E(p^2) - \Delta_0(p^2))M(p^2,q^2)(\Delta_E(q^2) -
\Delta_0(q^2))\ ,
\end{equation}
and $M(p^2,q^2)=p^2_{max} (C^{-1})(p^2,q^2)$ is related to the
covariance matrix $C(p^2,q^2)$.

\begin{figure}[!ht]
\begin{center}
\epsfxsize10cm\epsffile{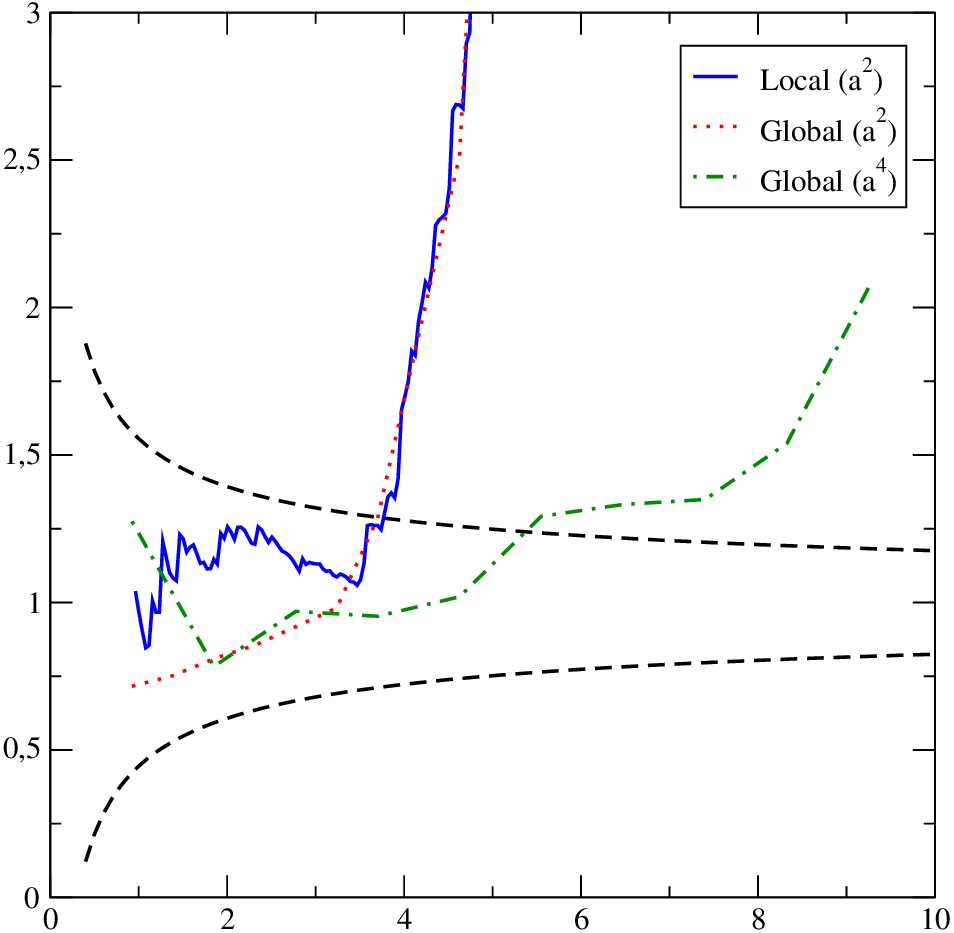}
\end{center}
\caption{\small{\it Evolution of the $\chi^2/d.o.f$ as a function of
    $p^2_{max}$ on a $32^4$ lattice ($a=m=1)$, for the local $a^2$ method
    (blue solid line), the global $a^2$ method (red dotted line) and the
    global $a^4$ method (green dash-dotted line). The smooth curves are
    the 95\% confidence levels lines - {\bf 1000 configurations}.}}
    \label{fig:chi2}
\end{figure}

With these considerations, we compute the $\chi^2/d.o.f.$ of a
zero-parameter fit of the extrapolated form factor to its known
value $\Delta_0(p^2)=1$ for all $p^2$. Figure~\ref{fig:chi2}
displays the evolution of the chisquare per degree of freedom as a
function of the fitting window $]5,n^2_{max}]$ on a $32^4$
lattice, for each extrapolation method. The local and global H4
methods which cure just $\mathcal{O}(a^2)$ artifacts are indeed
safe up to $p^2_{max}\approx \pi^2/4$.

For the range of lattice sizes and sample sizes considered in this
work, the global H4 method which takes into account
$\mathcal{O}(a^4)$ artifacts performs best. With such a method it
is possible to extend the range of validity of the extrapolation
towards the continuum up to $p^2\approx 5-6$, according to the
lattice size and at least down to the levels of statistical
accuracy studied here.

\section{Conclusion}

Table~\ref{tab:dem_vs_local} summarizes our findings. For each
lattice size, sample size and extrapolation method studied in this
work, the table displays the upper bound $p^2_{max}$ of the
momentum window $]0,a^2p^2_{max}]$ ($am=1$), inside which the
extrapolated dressing function $\Delta_E(p^2)/\Delta_0(p^2)$ is
consistent with 1 at a $\chi^2/d.o.f.=2$ level.

\begin{table}[ht]
\begin{center}
\begin{tabular}{|| c || c | c || c | c ||}
\hline
\hline
Lattice size & 32  & 32 & 64 & 64 \\
\hline
Sample size  & 100 & 1000 & 100 & 1000 \\
\hline
Democratic method  &  1.4 (2\%)       & 0.54 (0.9\%)& 1.8 (1.9\%) & 1.1 (0.6\%)  \\
\hline
Local $a^2$ method  & 6.3 (2\%)       & 4.2 (0.8\%) & 4.4 (1.4\%) & 3.4 (0.5\%) \\
\hline
Global $a^2$ method & 6.3 (0.7\%)     & 4.3 (0.3\%) & 4.0 (0.46\%) & 3.1 (0.15\%) \\
\hline
Global $a^4$ method & $\pi^2$ (0.9\%) & 9.2 (0.4\%) & $\pi^2$ (0.35\%) & 6.7 (0.12\%)\\
\hline
\hline
\end{tabular}
\end{center}
\caption{\small \it $p^2_{max}$ as a function of the lattice size
and the sample size for which $\chi^2/d.o.f.=2$. The statistical
error on the extrapolated dressing function is shown between
parentheses.} \label{tab:dem_vs_local}
\end{table}

The limits established in table \ref{tab:dem_vs_local} have been
obtained as described in section \ref{sec:H4}. They could be
improved by adding more terms to the Laurent's development, or
taking into account their perturbative form in the parametrization
of the artifacts.

Table~\ref{tab:dem_vs_local} is all what is needed to set up an H4
extrapolation towards the continuum. Our recommendations are the
following. If it is not required to push the extrapolation in
$a^2p^2$ above $\approx \pi^2/4$, then it is sufficient to use an
H4 method, either local or global, up to $\mathcal{O}(a^2)$
artifacts. On larger windows, the global H4 method at least up to
$\mathcal{O}(a^4)$ artifacts should be used. The precise tuning of
$p^2_{max}$ can be read off the table in each case.

The sample sizes used in this study are what is typically achieved
in lattice studies of two-point functions with $\mathcal{O}(1-10)$
GFlops computers. With sufficient time allocated on
$\mathcal{O}(1)$ Tflops computers, it would become possible to
increase the statistics by one or two orders of magnitude. Then
Table~\ref{tab:dem_vs_local} would no longer be accurate enough
and the analysis of this work would need to be repeated, including
the global H4 method up to $\mathcal{O}(a^6)$ artifacts in order
to keep the extrapolation windows as large. Let us emphasize that
such an analysis is straightforward to implement. With adequate
statistics, the global H4 extrapolation method can be
systematically improved.

A one or two order of magnitude increase of statistics would also
allow to apply the H4 extrapolation techniques to three-point
functions as well. Indeed, with a sample size around 1000
configurations, the discretization errors in such lattice
observables, although noticeable, are not large enough to be
separated from the statistical errors.  Three-point functions
depend on two momenta. It can be shown that there are now 14
\bfit{algebraically} independent symmetric invariants $\phi(p,q)$
under the action of the hypercubic group H(4), and among them, we
have the three $O(4)$ invariants
\begin{align*}
  \sum_{\mu} p^2_{\mu}\,,\quad\sum_{\mu} q^2_{\mu}\,,\quad
  \sum_{\mu} p_{\mu}q_{\mu}
\end{align*}
and 5 algebraically, and functionnally, independent invariants of
order $a^2$ which can be chosen as
\begin{align*}
    \sum_{\mu} p^4_{\mu}\,,\quad\sum_{\mu} q^4_{\mu}\,,\quad
    \sum_{\mu} p^2_{\mu}q^2_{\mu}\,,\quad\sum_{\mu} p^3_{\mu}q_{\mu}\,,\quad
    \sum_{\mu} p_{\mu}q^3_{\mu}
\end{align*}
Three-point form factors are usually measured only at special
kinematical configurations. Assuming again smoothness of the
lattice form factor with respect to these $\mathcal{O}(a^2)$
invariants, the global H4 extrapolation method could still be
attempted provided that enough lattice momenta and enough H4
orbits are included in the analysis.

A more straightforward application of the (hyper)cubic extrapolation
method is to asymmetric lattices $L^3\times T$ with spatial cubic
symmetry. Lattices with $T\gg L$ are produced in large scale
simulations of QCD with dynamical quarks at zero temperature, whereas
simulations of QCD at finite temperature require lattices with $T\ll
L$. For such lattices, the continuum limit can still be obtained
within each time slice by applying the techniques described in this
note to the cubic group $O_h$.

We want to end by pointing out that (hyper)cubic extrapolations
methods are not restricted to momentum space but can also be used
directly in spacetime. We will sketch one example for
illustration, the static potential.

Lattice artifacts show up in the static potential at short distances
and the standard recipe \cite{MICHAEL} to correct the artifacts is to
add to the functional form which fits the static potential a term
proportional to the difference $\delta G(R)$ between the lattice
one-gluon exchange expression and the continuum expression $1/R$. The
technique we advocate is rather to eliminate the cubic artifacts from
the raw data measured on the lattice.

Indeed the lattice potential extracted from the measurements of an
``off-axis'' Wilson loop connecting the origin to a point at distance
$R=\sqrt{x^2+y^2+z^2}$ can be expressed\footnote{at least for L-shaped
  loops.}, after averaging over the orbits of the cubic group $O_h$, as a
function of three invariants:
\begin{align*}
V_L(x,y,z) \equiv V_L(R^2,R^{[4]},R^{[6]})\,,\quad R^{[n]} = x^n+y^n+z^n
\end{align*}
An extrapolation towards the continuum can be performed with the
methods described in section~\ref{sec:artifact} by making the
smoothness assumption with respect to the invariants
$R^{[4]},\,R^{[6]}$.

\appendix
\subsection*{Acknowledgments}

We wish to thank our colleagues, Ph.~Boucaud, J.P.~Leroy, A. Le
Yaouanc, J.~Micheli, O.~P\`ene, J.~Rodr\'iguez-Quintero, who have
collaborated to the development of the H4 extrapolation method.
These calculations were performed at the IN2P3 computing center in
Lyon. F.S. is specially indebted to J. Carbonell and LPSC for
their warm hospitality.

\section{H(4) invariants}

A general polynomial of degree $N$ in the four components of the
momentum $p$ reads:
\begin{align*}
  P_N(p_1,p_2,p_3,p_4) = \sum_{n=0}^{N} \sum_{n_1+n_2+n_3+n_4=n} c_{n_1n_2n_3n_4}
  \,p_1^{n_1}p_2^{n_2}p_3^{n_3}p_4^{n_4}\ .
\end{align*}
But any polynomial function of $p$ which is invariant under the
action of $H(4)$ must be invariant under every permutation of the
components of $p$ and every reflection $p_{\mu}\rightarrow
-p_{\mu}$. In particular such a polynomial must be an even
function of each component $p_{\mu}$ and contain only symmetric
combinations of the components. As there are 4 components, we can
construct 4 symmetric combinations that are independent. They are
usually chosen as the elementary symmetric polynomials:
\begin{align*}
  \sigma_1 &= p_1^2+p_2^2+p_3^2+p_4^2 \\
  \sigma_2 &= p_1^2p_2^2+p_1^2p_3^2+p_1^2p_4^2+p_2^2p_3^2+p_2^2p_4^2+p^2_3p^2_4 \\
  \sigma_3 &= p^2_1p^2_2p^2_3+p^2_1p^2_2p^2_4+p^2_1p^2_3p^2_4+p^2_2p^2_3p^2_4 \\
  \sigma_4 &= p^2_1p^2_2p^2_3p^2_4
\end{align*}
Noticing that the variables $p_{\mu}^2$ are the roots of the polynomial
\begin{align*}
  Q(t) = t^4-\sigma_1 t^3+\sigma_2 t^2-\sigma_3 t +\sigma_4
\end{align*}
the invariant polynomial $P_N$ can be written, after recursive
substitution of all fourth powers of the $p^2_{\mu}$'s, as a
polynomial $\widetilde{P}_N$ in the four symmetric invariants:
\begin{align*}
  P_N(p_1,p_2,p_3,p_4) = \widetilde{P}_N(\sigma_1,\sigma_2,
  \sigma_3,\sigma_4)\ .
\end{align*}

We could have chosen other invariants to represent the polynomial,
as the power sums $p^{[n]} \equiv p_1^n+p_2^n+p_3^n+p_4^n$. They
can be indeed recovered from the symmetric invariants $\sigma_n$
via the recursive formulas:
\begin{align*}
   \sigma_1  &= p^{2} \\
 2\sigma_2 &= \sigma_1p^{2} - p^{[4]} \\
  3\sigma_3 &=  \sigma_2p^{2} - \sigma_1p^{[4]} + p^{[6]} \\
   4\sigma_4 &=  \sigma_3p^{2} - \sigma_2p^{[4]} +
   \sigma_1p^{[6]} - p^{[8]}\ .
\end{align*}
Thus, any polynomial on the four components of $p$ invariant under
the action of H(4) can be written as a polynomial in terms of the
power sums $p^{[n]}$. A complete, elegant proof can be found in
\cite{WEYL}.

\end{document}